\documentclass[a4paper]{article}
\usepackage{graphicx} 
\usepackage{subcaption}
\usepackage{color}
\usepackage{amsfonts}
\usepackage{amsmath}
\usepackage{mathtools}
\usepackage{physics}
\usepackage{dcolumn}
\usepackage{bm}
\usepackage{cancel}
\usepackage{tensor}
\usepackage{amsthm}
\usepackage{sectsty}
\usepackage[toc,page]{appendix}

\newtheorem{lemma}{Lemma}

\newtheorem{remark}{Remark}

\allowdisplaybreaks

\newcounter{mnotecount}

\newcommand{\mnotex}[1]
{\protect{\stepcounter{mnotecount}}$^{\mbox{\footnotesize $\bullet$\themnotecount}}$ 
\marginpar{
\raggedright\tiny\em
$\!\!\!\!\!\!\,\bullet$\themnotecount: #1} }

\def\bma{{\bm a}}
\def\bmb{{\bm b}}

\def\bme{{\bm e}}

\def\bmg{{\bm g}}
\def\bmh{{\bm h}}

\def\bmu{{\bm u}}

\def\bmomega{{\bm \omega}}

\def\bmA{{\bm A}}

\def\bmF{{\bm F}}

\def\bmP{{\bm P}}
\def\bmQ{{\bm Q}}

\def\bmT{{\bm T}}
\def\bmZ{{\bm Z}}

\begin{document}

\title{\textbf{On a field tensor for gravity and electromagnetism}}

\author{M Normann\footnote{\tt mikael.normann@usn.no}\\
{\em Faculty of Technology, Natural Sciences and Maritime Sciences}\\
{\em University of South-Eastern Norway} \\
{\em Porsgrunn, Norway}}


\date{\today}

\maketitle

\begin{abstract}
\noindent We show that a three rank Lanczos type tensor field is an appropriate choice to describe relativistic electromagnetic and gravitational effects. More precisely, we identify the irreducible field-decompositions of this tensor as gravitaional and electromagnetic fields. A set of divergence equations are proposed as field equations for the unified field.
\end{abstract}

\section{Introduction}
\noindent In the early to mid 1900 a number of articles were published on the unification of electromagnetism and gravitation. This program of unification has been put under the umbrella term Unified Field Theories (UFTs) --- see \cite{Goenner04} for a comprehensive review. But due to the remarkable achievment of Quantum Field theory in unifying the nuclear and electromagnetic forces, the UFT program has been replaced by the pursuit of a theory of Quantum Gravity. Since spinors are needed in the description of fermions \cite{ward_wells1990}, it is essential for a unified field theory to admit spinor structure in order to be a viable theory for the description of e.g. electrons. 
Geroch has shown in \cite{Geroch68} that it is a necessary and sufficient condition for a non-compact space time to admit a spinor structure if it carries a global field of orthonormal tetrads. The frame formalism also reflect the role of observers in physics, and is thus a natural formalism both in classical relativity and quantum field theory \cite{Wald1984}.
Furthermore, due to the nonlinearity of the Einstein equations, a metric distributional solution describing a point particle is not possible in general relativity \cite{Geroch1987}. We refer to \cite{Steinbauer2006} for a review of the use of distributions in general relativity. On the other hand, the Maxwell equations do admit a solution representing a charged point particle. In the present work we explore the possibility of a theory which both admits a spinor structure --- by employing a global tetrad field ---  and whose field equations are linear with respect to the sources and field tensor, in striking similarity with the Maxwell equations. We remark that we do not make use of the spinor structure in the present article. A proper investigation of the spinorial equations and detailed analysis of the spinor fields will be published elsewhere.
\section{Geometric considerations}
\label{section:GeometricCons}
Let $(\mathcal{M},\bmg)$ denote a spacetime represented by a 4-dimensional
manifold, $\mathcal{M}$, with a Lorentzian metric $\bmg$. The
motion of particles of some matter filling spacetime give rise to a
natural splitting by constructing frames comoving with the flow lines
of the particles. This has the further advantage that it does not
require a foliation of $\mathcal{M}$. We shall denote the tangent vector to the flow
lines as $\bmu$ satisfying
\[
\bmg(\bmu, \bmu) = -1.
\]
At each point $p \in \mathcal{M}$ the frame field
$\{\bme_{a}\}$ is such that
\[
\bmg(\bme_{a},\bme_{b}) = \eta_{a b},
\]
where $\eta_{ab}$ are the frame components of the Minkowski metric.
The frames $\{\bme_{a}\}$ give rise to a co-frame, $\{\mathbf{\omega}^{a}\}$ satisfying
\[
\langle{\bme_{a}, \mathbf{\bmomega}^{b}\rangle} = \delta_{a}{}^{b}.
\]
In the following all indices will be given in terms of the frame and co-frame unless otherwise stated. The metric tensor give rise to a natural connection $\mathbf{\nabla}$ such that $\mathbf{\nabla} \bmg = 0$,
which is the \textit{metric compatibility condition}. In terms of the frames, this condition takes the form
\begin{equation}
\label{metricComp}
\Gamma_{a}{}^{b}{}_{c} \eta_{bd} + \Gamma_{a}{}^{b}{}_{d} \eta_{bc} = 0,
\end{equation}
where the \textit{frame connection coefficients} are defined by the directional derivative along the direction of the frame indices
\[
\nabla_{a} \bme_{b} = \Gamma_{a}{}^{c}{}_{b} \bme_{c}, \qquad \nabla_{a} = \langle{\bme_{a}, \mathbf{\nabla}\rangle}.
\]
Thus, for a two rank tensor $\bm\Omega$ we have that the frame components of its derivative is given by,
\[
\nabla_{a}\Omega_{bc}= e_{c}[\Omega_{bc}] -\Gamma_{a}{}^{d}{b}\Omega_{dc}-\Gamma_{a}{}^{d}{c}\Omega_{bd}
\].
Furthermore, if the connection $\mathbf{\nabla}$ is \textit{torsion-free}, we have that
\begin{equation}
\label{torsionFree}
\Sigma_{a}{}^{c}{}_{b} = 0,
\end{equation}
where the frame components of the \textit{torsion tensor} are defined by
\[
\Sigma_{a}{}^{c}{}_{b} \bme_{c} = \left[\bme_{a}, \bme_{b}\right] + \left(\Gamma_{a}{}^{c}{}_{b} - \Gamma_{b}{}^{c}{}_{a}\right) \bme_{c}.
\]
The commutation of the connection may be expressed in terms of the \textit{Riemann curvature tensor} and the torsion tensor
\begin{eqnarray*}
&&\nabla_{[a}\nabla_{b]}v^{c} = R^{c}{}_{dab} v^{d} +\Sigma_{a}{}^{d}{}_{b}\nabla_{d}v^{c},\\
&&\nabla_{[a}\nabla_{b]}w_{c} = -R^{d}{}_{cab} w_{d} +\Sigma_{a}{}^{d}{}_{b}\nabla_{d} w_{c}\label{CommutatorRiemann}.
\end{eqnarray*}
The frame components of the Riemann curvature tensor is given by
\begin{equation}
R^c{}_{dab} = \partial_a\Gamma_b{}^c{}_d
   - \partial_b \Gamma_a{}^c{}_d +
   \Gamma_f{}^c{}_d(\Gamma_b{}^f{}_a -
   \Gamma_a{}^f{}_b) +
   \Gamma_b{}^f{}_d\Gamma_a{}^c{}_f -
   \Gamma_a{}^f{}_d\Gamma_b{}^c{}_f
   -\Sigma_a{}^f{}_b \Gamma_f{}^c{}_d \label{RiemannExpansion}
\end{equation}
---see \cite{Kroon2016} for details. The Riemann tensor has all the usual symmetries, and it satisfies the
\textit{Bianchi identity} for a general connection
\begin{eqnarray}
&& R^{d}{}_{[cab]} + \nabla_{[a}\Sigma_{b}{}^{d}{}_{c]} +
   \Sigma_{[a}{}^{e}{}_{b}\Sigma_{c]}{}^{d}{}_{e}=0, \label{1stBianchiId}\\
&& \nabla_{[a} R^{d}{}_{|e|bc]} + \Sigma_{[a}{}^{f}{}_{b} R^{d}{}_{|e|c]f} =0.\label{2ndBianchiId}
\end{eqnarray}
Furthermore, we recall that the Riemann tensor admits the \emph{irreducible decomposition}
\begin{eqnarray}
&& R^{c}{}_{dab} = C^{c}{}_{dab} + 2 (\delta^{c}{}_{[a}L_{b]d} -
   \eta_{d[a}L_{b]}{}^{c}), \label{RiemannDecomposition}
\end{eqnarray}
with $C^{c}{}_{dab}$ the components of the \emph{Weyl tensor} and 
\begin{equation}
S_{ab} \equiv R_{ab} -\frac{1}{6}R \eta_{ab}
\label{Definition:Schouten}
\end{equation}
denotes the components of the \emph{Schouten tensor}. The connection
$\mathbf{\nabla}$ is called the \textit{Levi-Civita connection} of $\bmg$ if it
satisfies \eqref{metricComp} and \eqref{torsionFree}. In what follows we
will assume the connection to be Levi-Civita.

\subsubsection*{A projection formalism}
At each point in the spacetime manifold $\mathcal{M}$ the flow lines give rise to a tangent
space which can be split into parts in the direction of $\bmu$
and those orthogonal. This means that without implying a foliation, we
may decompose every tensor defined at each point $p \in \mathcal{M}$
into its orthogonal and timelike part. This may be done by contracting
with $\mathbf{u}$ and the \textit{projector} defined as
\[
h_{a}{}^{b} \equiv \eta_{a}{}^{b} + u_{a}u^{b}, \qquad \bmu = u^{a}\mathbf{e}_{a}.
\]
Thus, a tensor $T_{ab}$ may be split into its time-like, mixed
and space-like parts given, respectively, by
\[
T_{00}= u^{a}u^{b}T_{ab}, \qquad T'_{0c}= u^{a}h^{b}{}_{c}T_{ab}, \qquad T'_{cd}= h^{a}{}_{c}h^{b}{}_{d}T_{ab},
\]
where $'$ denotes that the free indices left are spatial ---e.g. $T'_{a0} u^{a} = 0$. Decomposing $\mathbf{\nabla u}$ we
obtain
\begin{equation}
\label{Der4VelDecomp}
\nabla_{a} u^{b} = \chi_{a}{}^{b} - u_{a}a^{b},
\end{equation}
where $\chi_{a}{}^{b}$ and $a^{b}$ are the components of the
\textit{Weingarten tensor} and 4-acceleration, respectively, defined
by
\[
\chi_{a}{}^{b} \equiv h_{a}{}^{c}\nabla_{c} u^{b}, \qquad a^{b} \equiv u^{c}\nabla_{c}u^{b}.
\]
We split $\chi_{ab}$ into its symmetric, tracefree part and antisymmetric part --- i.e we have,
\[
\chi_{(ab)} - \frac{1}{3}h_{ab}\chi \equiv \sigma_{ab}, \qquad \chi_{[ab]} \equiv \omega_{\bma\bmb}.
\]
In the literature (e.g. see \cite{Wald1984} p.217) $\chi$,
$\sigma_{ab}$ and $\omega_{ab}$ is called,
respectively, the expansion, shear and the twist of the congruence with four velocity $\bmu$. The decomposition \eqref{Der4VelDecomp} now takes the form,
\begin{equation}
\label{Der4VelDecompFinal}
\nabla_{a} u^{b} = \sigma_{a}{}^{b} + \frac{1}{3} h_{a}{}^{b}\chi + \omega_{a}{}^{b}  - u_{a}a^{b}.
\end{equation}
The decomposition of the \textit{four volume} is
\[
\epsilon_{abcd} = -2\left(u_{[a}\epsilon_{b]cd}-\epsilon_{ab[c}u_{d]}\right), \qquad \epsilon_{bcd}=\epsilon_{abcd} u^{a}.
\]
Given a tensor $T_{abc}$ which is antisymmetric in its two last indices,
we may construct the \textit{electric} and \textit{magnetic} parts with respect to $\mathbf{u}$. In frame indices this is, respectively, defined by
\[
E_{cd} \equiv T_{abe} h_{c}{}^{a} h_{d}{}^{b} u^{e}, \qquad B_{cd} \equiv T^{\ast}{}_{abe} h_{c}{}^{a} h_{d}{}^{b} u^{e},
\]
where the \textit{Hodge dual operator}, denoted by ${}^{\ast}$, is defined
by 
\[
T^{\ast}{}_{abe} \equiv -\frac{1}{2}\epsilon^{mn}{}_{be} T_{amn},
\]
and has the property that
\[
T^{\ast \ast}{}_{abc} = -T_{abc}.
\]
Depending on the symmetries and rank of the tensor, the above
definition for electric and magnetic decomposition may vary
slightly. Central for our discussion is that $E_{ab}$ and $B_{ab}$
are spatial and symmetric.

\section{The field tensor}
\noindent We consider the rank three tensor $\bmZ$ (hereafter called the Z-tensor) with the following symmetries,
\[
Z_{[abc]} = 0 	,\qquad Z_{abc} = Z_{a[bc]}.
\]
It can be readily shown that the first symmetry property implies that
\begin{equation}
    Z_{cab} = 2Z_{[ba]c}.
\end{equation}
The Hodge dual of the Z-tensor $\bmZ^{\ast}$ is defined in the customary way by,
\[
Z^{\ast}{}_{abc}\equiv -\frac{1}{2}\epsilon_{bc}{}^{de}Z_{ade}.
\]
The frame fields $\bme_{a}$ provide a natural 1+3 decomposition of $\bmZ$ and $\bmZ^{\ast}$ into parts in the direction of and orthogonal to the flow $\bmu$. This is obtained by using the projector $\bmh$ as described in Section \ref{section:GeometricCons}.The decomposition read,
\begin{subequations}
\begin{eqnarray}
 &&Z_{abc} = -2\eta_{a[b} P_{c]} + \epsilon_{bc}{}^{d} \Phi_{ad} +
2u_{[b} \Psi_{c]a} -  \epsilon^{d}{}_{bc} u_{a} Q_{d} + 2\epsilon^{d}{}_{a[c} u_{b]} Q_{d}, \label{ZeldaElectricDecomp}\\
&&Z^{\ast}{}_{amn} = \epsilon_{mnb} u_{a} P^{b} - 2\epsilon_{ab[m} u_{n]}
P^{b} + 2 \Phi_{a[m}u_{n]} + \epsilon_{mnb} \Psi_{a}{}^{b} +
2\eta_{a[n} Q_{m]}, \label{ZeldaMagneticDecomp}
\end{eqnarray}
\end{subequations}
where we have defined,
\[
\Psi_{ab} \equiv Z_{(a'b')0}, 
 \qquad \Phi_{ab} \equiv Z^{\ast}_{(a'b')0},
\qquad P_{a} \equiv Z_{a00},
\qquad Q_{a} \equiv Z^{\ast}_{a00}.
\]
The tensors $\Psi_{ab}$ and $ \Phi_{ab}$ are by definition symmetric tensors defined on the orthogonal space of $\bmu$ ---i.e. one has that 
\[
\Psi_{ac} u^{a}=0, \qquad
\Phi_{ac} u^{a}=0.
\]
Furthermore, since  $ \epsilon_{abc}$, $\Psi_{ab}$ and $ \Phi_{ab}$ are spatial fields, it is readily shown that
\[
P_{0}=Q_{0}=0.
\]
The traces of the Z-tensor and its dual are,
\begin{align}
    & Z^{a}{}_{ba} = 3P_{b}+\Psi u_{b}, \qquad Z^{a}{}_{b}{}^{b} = 0, \label{TracesOfZelda}\\
    & Z^{\ast}{}^{a}{}_{ba} = 3Q_{b} - \Phi u_{b}, \qquad Z^{\ast}{}^{a}{}_{b}{}^{b} = 0 \label{TracesOfZeldaDual},
\end{align}
where,
\[
\Psi \equiv \Psi^{a}{}_{a}, \qquad \Phi \equiv \Phi^{a}{}_{a}.
\]
The first trace in \eqref{TracesOfZelda} implies that
\begin{equation}
     Z^{a}{}_{0a} = -\Psi,
\end{equation}
and the first trace in \eqref{TracesOfZeldaDual} together with the first symmetry property implies that
\begin{equation}
     Z^{\ast}{}^{a}{}_{0a} = \Phi=0.\label{TracePsi}
\end{equation}
\begin{lemma}
Let $\bmZ$ be a tensor of rank 3 with antisymmetry about two neighbouring indices. Then $\bmZ$ has the symmetry property
$Z_{[abc]}=0$ and the dual field $Z^{\ast}_{(a'b')0}$ has vanishing trace.\label{LemmaZ}
\end{lemma}
\noindent We make the further assumption that $\Psi=0$ --- i.e we have that
\[
 Z^{a}{}_{0a}=Z^{\ast}{}^{a}{}_{0a}=0.
\]
\begin{remark}
\em The assumption that $\Psi=0$ is motivated by the fact that we want to relate the fields $\bm\Psi$ and $\bm\Phi$ to the electric and magnetic part of the Weyl tensor. Observe that our assumption is a weaker constraint than the \textit{Lanczos algebraic gauge} --- e.g see \cite{Roberts1995}, \cite{Lanczos49},
\[
Z^{a}{}_{ba} =0.
\]
In fact, the Lanczos gauge violate our assumption that the fields $\bmP$ and $\bm\Psi$ represents pure electric and gravitational fields, respectively, and can thus not be related in such a way as this gauge implies --- see equation \eqref{TracesOfZelda}.
\end{remark}
\begin{remark}
    \em Observe that the absence of electric and magnetic fields is a necessary condition for the Z-tensor to be a a Cotton tensor.
\end{remark}
\section{Finding the field equations for the Z-tensor}
 In the theory we propose, both gravity and electromagnetism is represented in terms of a field on space time. The geometry of $\mathcal{M}$ will be given by the frame components, rather than the metric, and the connection coefficients as outlined in the introduction. Equations for the frame and the connection is given by the choice of propagation --- e.g. Fermi propagation --- and the definition of the Riemann and the torsion tensor. For more details on the geometric equations, the reader is referred to \cite{Normann2020}, \cite{Friedrich1998} and \cite{Pugliese2012}. In what follows we shall focus the discussion on the fields presented in the previous section --- i.e. $\bm\Psi$, $\bm\Phi$, $\bmP$ and $\bmQ$. These will be taken as the fundamental fields, from which we may construct the unified field tensor $\bmZ$. We thus seek a set of equations for $\bmZ$ which will reduce to the relativistic Maxwell equations in the limit of no gravitational field, and the Bianchi equations in the limit of no electromagnetic fields. We begin with the Maxwell equations. 
 \\
 
\noindent We observe that due to the symmetry of $\bmZ$ and $\bmZ^{\ast}$, it is natural to define the two rank anti-symmetric tensors $\bmF$ and $\bmF^{\ast}$ as follows,
\[
F_{ab}\equiv u^{a}Z_{abc},\qquad F^{\ast}{}_{ab}\equiv \dfrac{1}{2}\epsilon_{ab}{}^{mn}F_{mn} =  u^{a}Z^{\ast}{}_{abc}.
\]
Using the decomposition of the Z-tensor, it is readily shown that
\[
F_{ab} = u_{b} P_{a} -  u_{a} P_{b} + \epsilon_{abc} Q^{c},
\]
which is the right form of the Faraday tensor with $P_a$ and $Q_a$ as the electric and magnetic fields respectively. The Maxwell equations are then given by
\begin{subequations}
    \begin{align}
       \nabla^{b}F_{ab}&=j_c\\
        \nabla^{b}F^{\ast}{}_{ab}&=0,
    \end{align}
\end{subequations}
which may be formulated as evolution and constraint equations for the electric and magnetic fields --- i.e.
\begin{subequations}
    \begin{align}
    u^{a} h_{mb} \nabla_{a}E^{b} - \epsilon_{mab} \nabla^{b}B^{a} &= - a^{a} \epsilon_{mab} B^{b} +
J^{a} h_{ma} + E^{a} \chi_{am} -  E_{m} \chi^{a}{}_{a}.\label{EvolEqE}\\
\nabla_{a}E^{a} &= a^{a} E_{a} + u^{a} J_{a} -  \epsilon_{abc} B^{a},\label{DivE} \\
u^{b} h^{d}{}_{a} \nabla_{b}B^{a} &= a^{b} E^{a} \epsilon^{d}{}_{ba} +
B^{b} \chi_{b}{}^{d} -  B^{d} \chi^{b}{}_{b} -  \epsilon^{d}{}_{ba} \nabla^{a}E^{b}
\chi^{bc},\label{EvolEqB}\\
\nabla_{b}B^{b} &= a^{b} B_{b} + E^{b} \epsilon_{bac} \chi^{ac}.\label{DivB}
    \end{align}
\end{subequations}

\noindent We now turn to consider equations for the gravitational field. It is customary to here study solutions to the Einstein field equations --- i.e 
\begin{equation}
R_{ab}-\dfrac{1}{2}Rg_{ab} = \tau_{ab}.\label{EFE}
\end{equation}
But as we are seeking a theory where the geometry is given by the frame components and the gravitational field is represented by the irreducible components of the Weyl tensor, we will use the Bianchi identity \eqref{2ndBianchiId} as field equations. In this formalism the Einstein equations
takes on the form of constraint equations --- see equation \eqref{SchoutenToEnergyMomentum}. Thus, the unknowns for the gravitational field will be the electric $E_{ab}$ and magnetic $B_{ab}$ part of the Weyl tensor --- i.e we consider the equations
\begin{subequations}
\begin{align}
u^{a} h_{m}{}^{c} h_{n}{}^{d} \nabla_{a}E_{cd} +\epsilon_{mdc} h_{n}{}^{a}
\nabla^{d}B_{a}{}^{c} &= - 2a^{a}
B_{(m}{}^{c}\epsilon_{n)ac} - 2E_{mn} \chi^{a}{}_{a} -  E_{ac} h_{mn} \chi^{ac} \nonumber\\
&+ 2 E_{na} \chi^{a}{}_{m} + E_{ma} \chi_{n}{}^{a} -  \tfrac{1}{2} u^{a}h_{m}{}^{c} h_{n}{}^{d} \nabla_{a}S_{cd}\nonumber\\
\qquad & + \tfrac{1}{2} u^{a}
h_{m}{}^{c} h_{n}{}^{d} \nabla_{d}S_{ac}\label{EvolWeylElectric}\\
\nabla_{a}E_{d}{}^{a} &= a^{a} E_{da} + E_{ac} u_{d} \chi^{ac} - \epsilon_{dcf} B_{a}{}^{f} \chi^{ac} - \epsilon_{acf} B_{d}{}^{f}\nonumber \\
& \chi^{ac} -  \tfrac{1}{2} u^{a} u^{c} \nabla_{c}S_{da} + \tfrac{1}{2} u^{a} u^{c} \nabla_{d}S_{ac},\label{DivWeylElectric}\\
u^{a} h_{l}{}^{c} h_{n}{}^{d} \nabla_{a}B_{cd} - \epsilon_{dc(n}
h_{l)}{}^{a} \nabla^{d}E_{a}{}^{c}&= 2a^{a} E_{(n}{}^{c}
\epsilon_{l)ac} - 2 B_{ln}
\chi^{a}{}_{a} -  B_{ac} h_{ln} \chi^{ac}\nonumber\\
& +  2\chi^{a}{}_{(l}B_{n)a} 
 + B_{a(n} \chi_{l)}{}^{a} + \tfrac{1}{2} \epsilon_{cd(n}
h_{l)}{}^{a} \nabla^{d}S_{a}{}^{c},\label{EvolWeylMagnetic}\\
h_{n}{}^{a} \nabla_{c}B_{a}{}^{c}& = a^{a} B_{na} -  E_{c}{}^{d}
\epsilon_{nad} \chi^{ac} + 2 E_{a}{}^{d} \epsilon_{ncd} \chi^{ac}\nonumber\\
&+ \tfrac{1}{2} \epsilon_{ncd} u^{a} \nabla^{d}S_{a}{}^{c}\label{DivWeylMagnetic}
\end{align}
\end{subequations}
\noindent where $S_{ab}$ is the Schouten tensor and defined in the customary way --- see equation \eqref{Definition:Schouten}.
   If the Einstein equations are assumed, then the Schouten tensor is related to the Energy-momentum tensor $\tau_{ab}$ according to
    \begin{equation}
    S_{ab} = \tau_{ab}  -  \tfrac{1}{3} \tau^{c}{}_{c} \ g_{ab}.\label{SchoutenToEnergyMomentum}
    \end{equation}
Thus, a solution $(E_{ab}, B_{ab})$ of the evolution equations \eqref{EvolWeylElectric} and \eqref{EvolWeylMagnetic}, satisfying the constraint equaitons \eqref{DivWeylElectric} and \eqref{DivWeylMagnetic}, together with equation \eqref{SchoutenToEnergyMomentum} is equivalent to a metric solution of the Einstein equations \eqref{EFE} for a given energy momentum tensor $\tau_{ab}$ --- again the reader is referred to \cite{Normann2016} for more details.
\\

\noindent Observe that $Z_{abc}$ contains all the fields necessary for a description of both gravity and electromagnetism. That is, the spatial fields ($P_a, Q_a, \Psi_{ab}, \Phi_{ab}$) has the correct rank, trace and symmetry to represent ($E_a, B_a, E_{ab}, B_{ab}$), respectively. The strategy to find the correct field equations for the unified field tensor $Z_{abc}$ is to compare the proposed equations so that they reduce to the Maxwell equations and Bianchi equations in the case of no gravity and electromagnetism, respectively. That is, we must construct the equations such that  $\bm\Psi$ and $\bm\Phi$ will be a solution of equations \eqref{EvolWeylElectric} - \eqref{DivWeylMagnetic} when $P_a = Q_a = 0$. Similarly, $P_a, Q_a$ is requiered to be a solution of equations \eqref{EvolEqE} - \eqref{DivB} in the limit of $\Psi_{ab}=\Phi_{ab}=0$. 
\\

\noindent Due to the form of the decomposition of $\bmZ$, we propose field equations on the form,
\begin{subequations}
    \begin{align}
    \nabla^{b}Z_{abc}&=T_{ac},\label{FieldEq1}\\
    \nabla^{b}Z^{\ast}{}_{abc}&=A_{ac}.\label{FieldEq2}
    \end{align}
\end{subequations}
Note that as a consequence of the antisymmetry in the Z-tensor and the symmetry of the Ricci tensor, it follows that
\begin{equation}
    \nabla^c T_{ac} = \nabla^c A_{ac} = 0. 
\end{equation}
\begin{remark}
    \em For generality we shall not impose symmetry about the indices $\{a,c\}$, so as to make $\bm T$ and $\bm A$ symmetric tensors. But strictly speaking such an assumption should be made in order to study the equations on the form that most resembles the Bianchi equations and the relativistic Maxwell equations.
    Furthermore, this will make the tensors $\bm A$ and $\bm T$ divergence free.
\end{remark}

\noindent In what follows, we will show that there exists tensors $A_{ab}$ and $T_{ab}$ such that the proposed field equations encompass the relativistic Maxwell equations as well as the Bianchi equations.
Recall that any tensor $\bmT$ may be decomposed in parts orthogonal and parallel to the four velocity $\bmu$ according to,
\[
T_{ab}=T_{a'b'} + T_{a'0}u_b + T_{0 b'}u_a + T_{0 0}u_a u_b.
\] 
We consider first the spatial components of the field equations --- i.e 
\begin{subequations}
    \begin{align}
       h_{m}{}^{a}h_{n}{}^{c}\nabla^{b}Z_{abc}&=T_{m'n'}\label{DivZ1}\\
       h_{m}{}^{a}h_{n}{}^{c}\nabla^{b}Z{}^{\ast}_{abc}&=A_{m'n'}\label{DivZ2}.
    \end{align}
\end{subequations}
Using the decomposition of $Z_{abc}$ and $Z^{\ast}{}_{abc}$ \eqref{DivZ1} and \eqref{DivZ2} are equivalent to,
\begin{subequations}
    \begin{align}
       u^{a} h_{m}{}^{b} h_{n}{}^{c} \nabla_{a}\Psi_{bc} + \epsilon_{mbc}
h_{n}{}^{a} \nabla^{c}\Phi_{a}{}^{b} &= - a_{n} P_{m} +
a^{a} \epsilon_{nab} \Phi_{m}{}^{b} + a^{a} \epsilon_{mab}
\Phi_{n}{}^{b} - 2 \Psi_{mn} \chi^{a}{}_{a}\\
&  -  h_{mn} \Psi_{ab}
\chi^{ab} + 2 \Psi_{na} \chi^{a}{}_{m} + \epsilon_{mna} Q^{a}\chi^{b}{}_{b} -  \epsilon_{nab} Q^{a} \chi^{b}{}_{m} \nonumber\\
& + \epsilon_{mab} Q^{a} \chi^{b}{}_{n} + \Psi_{ma} \chi_{n}{}^{a} -
h_{m}{}^{b} h_{n}{}^{c} P^{a} \nabla_{a}h_{bc} -  h_{mn} \nabla_{a}P^{a} \nonumber\\
& + \epsilon_{mnb} u^{a} \nabla_{a}Q^{b} -  \tfrac{1}{2}
u^{a} h_{m}{}^{b} h_{n}{}^{c} \nabla_{a}S_{bc} + h_{n}{}^{a} P_{m} \nabla_{b}h_{a}{}^{b}\nonumber\\
& + h_{ma} h_{nb} \nabla^{b}P^{a} + \tfrac{1}{2}
u^{a} h_{m}{}^{b} h_{n}{}^{c} \nabla_{c}S_{ab},\nonumber \\
u^{a} h_{m}{}^{b} h_{n}{}^{c} \nabla_{a}\Phi_{bc} -\epsilon_{mbc} h_{n}{}^{a} \nabla^{c}\Psi_{a}{}^{b} &= -2 a^{a} \epsilon_{mab} \Psi_{n}{}^{b} + a_{n} Q_{m} - 2 \Phi_{mn} \chi^{a}{}_{a} -  h_{mn} \Phi_{ab} \chi^{ab}\\
& + 2\Phi_{ma} \chi^{a}{}_{n} + \epsilon_{mna} P^{a} \chi^{b}{}_{b} -  \epsilon_{nab} P^{a} \chi^{b}{}_{m} + \epsilon_{mab} P^{a} \chi^{b}{}_{n} \nonumber \\
& + \Phi_{na} \chi_{m}{}^{a} + h_{m}{}^{b}
h_{n}{}^{c} Q^{a} \nabla_{a}h_{bc} + \epsilon_{mnb} u^{a}\nabla_{a}P^{b} + \epsilon_{mnb}\nabla_{a}\Psi^{ab}\nonumber \\
&+ h_{mn}\nabla_{a}Q^{a} -  h_{n}{}^{a} Q_{m}\nabla_{b}h_{a}{}^{b} - h_{ma} h_{nb}\nabla^{b}Q^{a} -  \tfrac{1}{2}\epsilon_{mbc} h_{n}{}^{a}\nabla^{c}S_{a}{}^{b}\nonumber
    \end{align}
\end{subequations}
where we have defined,
\begin{subequations}
    \begin{align}
    h_{mc} h_{na} T^{ac} & \equiv a^{a} \epsilon_{nac} \Phi_{m}{}^{c} - \Psi_{mn} \chi^{a}{}_{a} -  h_{mn} \Psi_{ac} \chi^{ac} + \Psi_{na} \chi^{a}{}_{m}\label{Tequation}\\ 
    &+ \Psi_{ma} \chi_{n}{}^{a} -  \tfrac{1}{2} u^{a} h_{m}{}^{c} h_{n}{}^{d} \nabla_{a}S_{cd} + \tfrac{1}{2} u^{a} h_{m}{}^{c} h_{n}{}^{d} \nabla_{d}S_{ac} \nonumber \\
   A^{ac} h_{mc} h_{na} & \equiv a^{a} \epsilon_{mac} \Psi_{n}{}^{c} + \Phi_{mn} \chi^{a}{}_{a} + h_{mn} \Phi_{ac} \chi^{ac} + \Phi_{na} \chi^{a}{}_{m} - 2 \Phi_{ma} \chi^{a}{}_{n} \label{Aequation}\\
   & -  \Phi_{na} \chi_{m}{}^{a} -  \epsilon_{mnc} \nabla_{a}\Psi^{ac} + \tfrac{1}{2} \epsilon_{mcd} h_{n}{}^{a} \nabla^{d}S_{a}{}^{c}\nonumber 
    \end{align}
\end{subequations}
Thus the spatial components of $\bmT$ and $\bmA$ are determined by the assumption that in the absence of electromagnetic fields, equations \eqref{DivZ1} and \eqref{DivZ2} reduce to equations \eqref{EvolWeylElectric} and \eqref{DivWeylMagnetic}, respectively, under the identifications $\Psi_{ab}=E_{ab}$ and $\Phi_{ab} = -B_{ab}$.
 Next we consider mixed components. $T_{a'0}$ and $A_{a'0}$ are obtained by comparing with the Bianchi constraint equations. We consider the equations
\begin{subequations}
\begin{align}
    h^{a}{}_{d}u^{c}\nabla^{c}Z_{abc} &= h^{a}{}_{d}u^{c}T_{ac},\label{SpatialDivZ}\\
    h^{a}{}_{d}u^{c}\nabla^{b}Z{}^{\ast}_{abc} &= h^{a}{}_{d}u^{c}A_{ac}.\label{SpatialDivZDual}
    \end{align}
\end{subequations}
Again, using the decomposition of the Z tensor and its dual, \eqref{SpatialDivZ} and \eqref{SpatialDivZDual} are equivalent to
   \begin{subequations}
    \begin{align}
    h_{n}{}^{a} \nabla_{b}\Psi_{a}{}^{b} &= a^{a} \Psi_{na} + \epsilon_{nbc} \Phi_{a}{}^{c}\chi^{ab} + \epsilon_{abc} \Phi_{n}{}^{c} \chi^{ab} -  P^{a} \chi_{na}\\
    &-  \tfrac{1}{2} u^{a} u^{b} h_{n}{}^{c} \nabla_{b}S_{ac} + \epsilon_{nab} \nabla^{b}Q^{a} + \tfrac{1}{2} u^{a} u^{b} h_{n}{}^{c} \nabla_{c}S_{ab},\nonumber\\
h_{n}{}^{a} \nabla_{b}\Phi_{a}{}^{b} &= a^{a} \Phi_{na} - 2\epsilon_{nbc} \Psi_{a}{}^{c} \chi^{ab} + \epsilon_{nac} \Psi_{b}{}^{c} \chi^{ab} + Q^{a}\chi_{na}\\
& +\epsilon_{nab}\nabla^{b}P^{a} - \tfrac{1}{2}\epsilon_{nbc} u^{a} \nabla^{c}S_{a}{}^{b},
 \end{align}
\end{subequations}
where we have defined
\begin{subequations}
    \begin{align}
    h^{b}{}_{d}u^{a} T_{ba} &\equiv \epsilon_{dcf} \Phi_{a}{}^{f} \chi^{ac} -  \tfrac{1}{2} u^{a} u^{c} \nabla_{c}S_{da} + \tfrac{1}{2} u^{a} u^{c} \nabla_{d}S_{ac}, \label{Ta0}\\
h^{b}{}_{d} u_{a}A_{n}{}^{a} & \equiv  2 \epsilon_{ncd}
\Psi_{a}{}^{d} \chi^{ac} -  \epsilon_{nad} \Psi_{c}{}^{d} \chi^{ac} -
 \epsilon_{acd} \Psi_{n}{}^{d} \chi^{ac} \label{Aa0} \\
&  + \tfrac{1}{2}
\epsilon_{ncd} u^{a} \nabla^{d}S_{a}{}^{c}.\nonumber
 \end{align}
\end{subequations}
The other mixed components $T_{0b'}$ and $A_{0b'}$ are determined by comparing with the relativistic Maxwell equations in the limit of no gravitational fields:
 \begin{subequations}
    \begin{align}
    h^{c}{}_{d}u^{a}\nabla^{b}Z_{abc} &  = h^{c}{}_{d}u^{a}T_{ac},\label{ConstraintEq1}\\
    h^{c}{}_{d}u^{a}\nabla^{b}Z{}^{\ast}_{abc} &  = h^{c}{}_{d}u^{a}A_{ac}.\label{ConstraintEq2}
 \end{align}
 \end{subequations}
 The decomposed equations are given by
 \begin{subequations}
     \begin{align}
         u^{a} h_{mb} \nabla_{a}P^{b} - \epsilon_{mab} \nabla^{b}Q^{a} &= J^{a} h_{ma} -  a^{a} \Psi_{ma} -  a^{a} \epsilon_{mab} Q^{b} + P^{a} \chi_{am} \\
         &-  P_{m} \chi^{a}{}_{a} + \epsilon_{mac} \Phi_{b}{}^{c} \chi^{ab},\nonumber\\
         u^{a} h_{mb} \nabla_{a}Q^{b} +  \epsilon_{mab} \nabla^{b}P^{a} &= a^{a} \epsilon_{mab} P^{b} + a^{a} \Phi_{ma} + Q^{a} \chi_{am} -  Q_{m} \chi^{a}{}_{a} \nonumber\\
         &+\epsilon_{mac} \Psi_{b}{}^{c} \chi^{ab},
     \end{align}
 \end{subequations}
 where,
 \begin{subequations}
    \begin{align}
    u^{a} h_{mb} T_{a}{}^{b} &= - J^{a} h_{ma} + a^{a} \epsilon_{mab} Q^{b} -  P^{a} \chi_{am} + P_{m} \chi^{a}{}_{a}, \label{T0b}\\
A^{ba} u_{b} h^{d}{}_{a} &= - a^{b} \epsilon^{d}{}_{ba} P^{a} -  Q^{b} \
\chi_{b}{}^{d} + Q^{d} \chi^{b}{}_{b}.\label{A0b}
 \end{align}
\end{subequations}
Finally, we find $T_{00}$ and $A_{00}$ by using the electromagnetic divergence equations. Thus, we consider the equations,
\begin{subequations}
    \begin{align}
    u^{c}u^{a}\nabla^{b}Z_{abc} &  = u^{c}u^{a}T_{ac},\\
    u^{c}u^{a}\nabla^{b}Z{}^{\ast}_{abc} &  = u^{c}u^{a}A_{ac}.
 \end{align}
 \end{subequations}
 Again, by the decomposition of the Z-tensor and its dual, these are equivalent to the divergence equations
\begin{subequations}
     \begin{align}
         \nabla_{a}P^{a} &= u^{a} J_{a} + a^{a} P_{a} -  \Psi_{ab} \chi^{ab} - \epsilon_{abc} Q^{a} \chi^{bc},\\
         \nabla_{a}Q^{a} &= a^{a} Q_{a} + \Phi_{ab} \chi^{ab} + \epsilon_{abc} P^{a} \chi^{bc}.
     \end{align}
 \end{subequations}
As before, we have in the above equations defined,
 \begin{subequations}
     \begin{align}
         u^{a} u^{b} T_{ab} &= - u^{a} J_{a} + \epsilon_{abc} Q^{a} \chi^{bc}\\
         A^{ba} u_{a} u_{b} &= - \epsilon_{bac} P^{b} \chi^{ac}.
     \end{align}
 \end{subequations}
 



We have thereby shown that if $\bm A$ and $\bmT$ are given by,
\begin{subequations}
\begin{align}
    T_{ab} &= - u_{a} u_{b} u^{m} J_{m} -  u_{a} J^{m} h_{bm} + a^{m} \epsilon_{amn} \Phi_{b}{}^{n} + a^{m} \epsilon_{bmn} u_{a} Q^{n} \nonumber\\
    & \quad + \Psi_{bm} \chi_{a}{}^{m} -  u_{a} P^{m} \chi_{mb} + \Psi_{am} \chi^{m}{}_{b} + u_{a} P_{b} \chi^{m}{}_{m} -  \Psi_{ab}
\chi^{m}{}_{m} \nonumber\\
& \quad + \epsilon_{anc} u_{b} \Phi_{m}{}^{c} \chi^{mn} -  h_{ab} \Psi_{mn} \chi^{mn} + \epsilon_{mnc} u_{a} u_{b} Q^{m} \chi^{nc} + \tfrac{1}{2} u_{b} u^{m} u^{n} h_{a}{}^{c} \nabla_{c}S_{mn} \nonumber\\
& \quad -  \tfrac{1}{2} u^{m} h_{a}{}^{n} h_{b}{}^{c} \nabla_{m}S_{nc} -  \tfrac{1}{2} u_{b} u^{m} u^{n} h_{a}{}^{c} \nabla_{n}S_{mc} + \tfrac{1}{2} u^{m} h_{a}{}^{n} h_{b}{}^{c} \nabla_{n}S_{mc},\\
& \nonumber\\
A_{ab} &= - a^{m} \epsilon_{bmn} u_{a} P^{n} + a^{m} \epsilon_{bmn}
\Psi_{a}{}^{n} -  \Phi_{am} \chi_{b}{}^{m} -  u_{a} Q^{m} \chi_{mb} \nonumber\\
& \quad -2 \Phi_{bm} \chi^{m}{}_{a} + \Phi_{am} \chi^{m}{}_{b} + \Phi_{ab}
\chi^{m}{}_{m} + u_{a} Q_{b} \chi^{m}{}_{m} \nonumber\\
& \quad + h_{ab} \Phi_{mn}
\chi^{mn} -  \epsilon_{mnc} u_{b} \Psi_{a}{}^{c} \chi^{mn} + 2
\epsilon_{anc} u_{b} \Psi_{m}{}^{c} \chi^{mn} -  \epsilon_{amc} u_{b}
\Psi_{n}{}^{c} \chi^{mn} \nonumber\\
& \quad -  \epsilon_{mnc} u_{a} u_{b} P^{m}
\chi^{nc} + \tfrac{1}{2} \epsilon_{anc} u_{b} u^{m} \nabla^{c}S_{m}{}^{n} + \tfrac{1}{2} \epsilon_{bnc} h_{a}{}^{m} \nabla^{c}S_{m}{}^{n} + \epsilon_{abn} \nabla_{m}\Psi^{mn},\label{ADefinition}
\end{align}
\end{subequations}
then there exists a solution of the field  equations \eqref{FieldEq1} and \eqref{FieldEq2}, which are also solutions to the Bianchi equations and the relativistic Maxwell equations under appropriate limits. Then, they will also be a solution of the Einstein equations if the constraint equation \eqref{SchoutenToEnergyMomentum} is imposed. Observe that the divergence of $\bm\Psi$ in equation \eqref{ADefinition} will vanish if symmetry of $\bmA$ and $\bmT$ is assumed. Since the tensors $\bmT$ and $\bmA$ act as sources for the field tensor, it is worth mentioning that it is perturbations of the four velocity and the Schouten tensor which is responsible for a non-vanishing source. That is, a solution $(\bm e_a,\bm\Gamma)$ to the geometric equations, determines a solution to the field equations \eqref{FieldEq1} and \eqref{FieldEq2}. Wee see in this formalism that the Einstein equation is only a particular solution for a specific choice of geometry --- i.e. the Ricci tensor and scalar takes a specific form according to the matter distribution. In the theory proposed here, the perturbations of the Schouten tensor and frame components creates a matter distribution in space time which in turn produces gravitational and electromagnetic fields.
\section{Discussion}
\noindent It has been shown that it is possible to interpret $\bm\Psi$, $\bm\Phi$, $\bmP$ and $\bmQ$ as the gravitational and electromagnetic fields, respectively. Although there remains work to be done on the interpretations of these equations as well as the relation to the Einstein-Maxwell equations, we have shown that the tensor $\bmZ$ can be considered a viable candidate for 
a unified field theory where the tensors $\bm T$ and $\bm A$ are the sources --- see equations \eqref{FieldEq1} and \eqref{FieldEq2} --- and the field equations are first order divergence equations, in striking similarity to the Maxwell equations. Due to the existence of a global tetrad field it is natural to consider the spinorial formulation of the equations. This would be of interest for a possible quantum description as well as a more lucid interpretation of the equations. Another interesting further study would be the existence of solutions representing a charged point particle. The similarity of the equations with the Maxwell equations may suggest that such a solution exists and makes sense. But observe that although the field equations resembles the form of the Maxwell equations, there are derivatives in the sources which may create complications. 
\bibliography{references}
\bibliographystyle{abbrv}
\end{document}